\documentclass[pra,twocolumn,aps,amssymb,footinbib]{revtex4-1}
  \usepackage[dvips]{color}

\usepackage{amsfonts}
\usepackage{amsmath}
\usepackage{amssymb}
\usepackage{graphicx}%
\providecommand{\U}[1]{\protect\rule{.1in}{.1in}}
\begin{document}

\preprint{APS/123-QED}

\title{Topological Flat Bands from Dipolar Spin Systems}

\author{N. Y. Yao$^{1\dagger*}$, C. R. Laumann$^{1,2\dagger}$, A. V. Gorshkov$^{3\dagger}$, S. D. Bennett$^{1}$, E. Demler$^{1}$, P. Zoller$^{4}$, M. D. Lukin$^{1}$}

%\author{N. Y. Yao$^{1\dagger*}$}
%\affiliation{Department of Physics, Harvard University, Cambridge, MA 02138, U.S.A.}
%\author{C. R. Laumann$^{1,2\dagger*}$}
%\affiliation{Department of Physics, Harvard University, Cambridge, MA 02138, U.S.A.}
%\affiliation{ITAMP, Harvard University, Cambridge, MA 02138, U.S.A.}
%\author{A. V. Gorshkov}
%\affiliation{Institute for Quantum Information and Matter, California Institute of Technology, Pasadena, CA 91125, U.S.A.}
%\author{S. D. Bennett}
%\affiliation{Department of Physics, Harvard University, Cambridge, MA 02138, U.S.A.}
%\author{E. Demler}
%\affiliation{Department of Physics, Harvard University, Cambridge, MA 02138, U.S.A.}
%\author{P. Zoller}
%\affiliation{Institute for Quantum Optics and Quantum Information of the Austrian Academy of Sciences, A-6020 Innsbruck, Austria}
%\author{M. D. Lukin}
%\affiliation{Department of Physics, Harvard University, Cambridge, MA 02138, U.S.A.}

\affiliation{$^{1}$Physics Department, Harvard University, Cambridge, MA 02138, U.S.A.}
\affiliation{$^{2}$ITAMP, Harvard-Smithsonian Center for Astrophysics, Cambridge, MA 02138, U.S.A.}
\affiliation{$^{3}$Institute for Quantum Information and Matter, California Institute of Technology, Pasadena, CA 91125, U.S.A.}
\affiliation{$^{4}$Institute for Quantum Optics and Quantum Information of the Austrian Academy of Sciences, A-6020 Innsbruck, Austria}
\affiliation{$^{\dagger}$These authors contributed equally to this work}
\affiliation{$^{*}$e-mail: nyao@fas.harvard.edu}
\date{\today}% It is always \today, today,
\begin{abstract}
We propose and analyze a physical system that naturally admits two-dimensional topological nearly flat bands. 
Our approach utilizes an array of three-level dipoles (effective $S=1$ spins) driven by inhomogeneous electromagnetic fields. 
The dipolar interactions produce arbitrary uniform background gauge fields for an effective collection of conserved hardcore bosons, namely, the dressed spin-flips. These gauge fields result in topological band structures, whose bandgap can be  larger than the corresponding bandwidth. 
%
%We show that dipolar interactions can generate arbitrary uniform gauge fields in such a system, resulting in topological band structures; we demonstrate that the associated bandgap can be  larger than the corresponding bandwidth. 
 %
%Specifically, we consider a two-dimensional lattice of driven, tilted dipoles.  
%
%The dipolar interactions produce arbitrary uniform background gauge fields for an effective collection of conserved hardcore bosons, namely, the dressed spin-flips. These gauge fields result in , whose bandgap. 
%
%
Exact diagonalization of the full interacting Hamiltonian at half-filling reveals the existence of superfluid, crystalline, and supersolid phases. 
An experimental realization using either ultra-cold polar molecules or spins in the solid state is considered.

\end{abstract}

\pacs{73.43.Cd, 05.30.Jp, 37.10.Jk, 71.10.Fd}
\keywords{ultracold atoms, polar molecules, gauge fields, flat bands, superfluid, supersolid, dipolar interactions}
\maketitle

% TODO Make the single-particle versus many-body sectioning more explicit

Single-particle flat bands, where kinetic energy is quenched relative to the scale of interactions, are being actively explored in the quest for novel strongly correlated phases of matter \cite{Roy12, Regnault11, Moller12, Sheng11,  Parameswaran11, Liu12,Wang11,McGreevy12}. 
Prompted by the analogy to Landau levels, recent efforts have focused on \emph{topological} flat bands (TFB) -- lattice models in which the band-structure also harbors a non-trivial Chern invariant. Seminal recent work has highlighted that certain classes of highly-engineered two-dimensional tight binding models can indeed exhibit topological nearly flat bands \cite{Sun11, Tang11,Neupert11, Trescher12, Yang12, Wang12}. However, the identification of a physical system whose microscopics naturally admit TFB remains an outstanding challenge.

In this Letter, we demonstrate the emergence of synthetic gauge fields for an ensemble of interacting hardcore bosons --- the effective spin-flips of pinned, three-level dipoles in a  two-dimensional lattice.
%. Our approach involves a two-dimensional lattice of pinned, three-level dipoles and 
Underlying these gauge fields are two key ingredients: spatially varying, elliptically-polarized external  (microwave or optical) fields  break time-reversal symmetry, while anisotropic dipolar interactions  induce orientation-dependent phases onto the hopping hardcore bosons. 
The combination of these effects naturally produces nontrivial Chern numbers in the band structure and, when tuned appropriately, results in the emergence of flat bands due to hopping interference. While we observe a variety of non-topological correlated many-body states here (ranging conventional crystals to supersolids), interacting particles living in such a flat-band-kinetic environment are also leading candidates for the realization of fractional Chern insulators \cite{Roy12, Regnault11, Moller12, Sheng11,  Parameswaran11, Liu12,Wang11,McGreevy12}.  Our proposal describes a natural framework in which ultra-cold molecules may be used to probe the exotic features of such interacting topological insulators.

\begin{figure}
\centering
\includegraphics[width=3.4in]{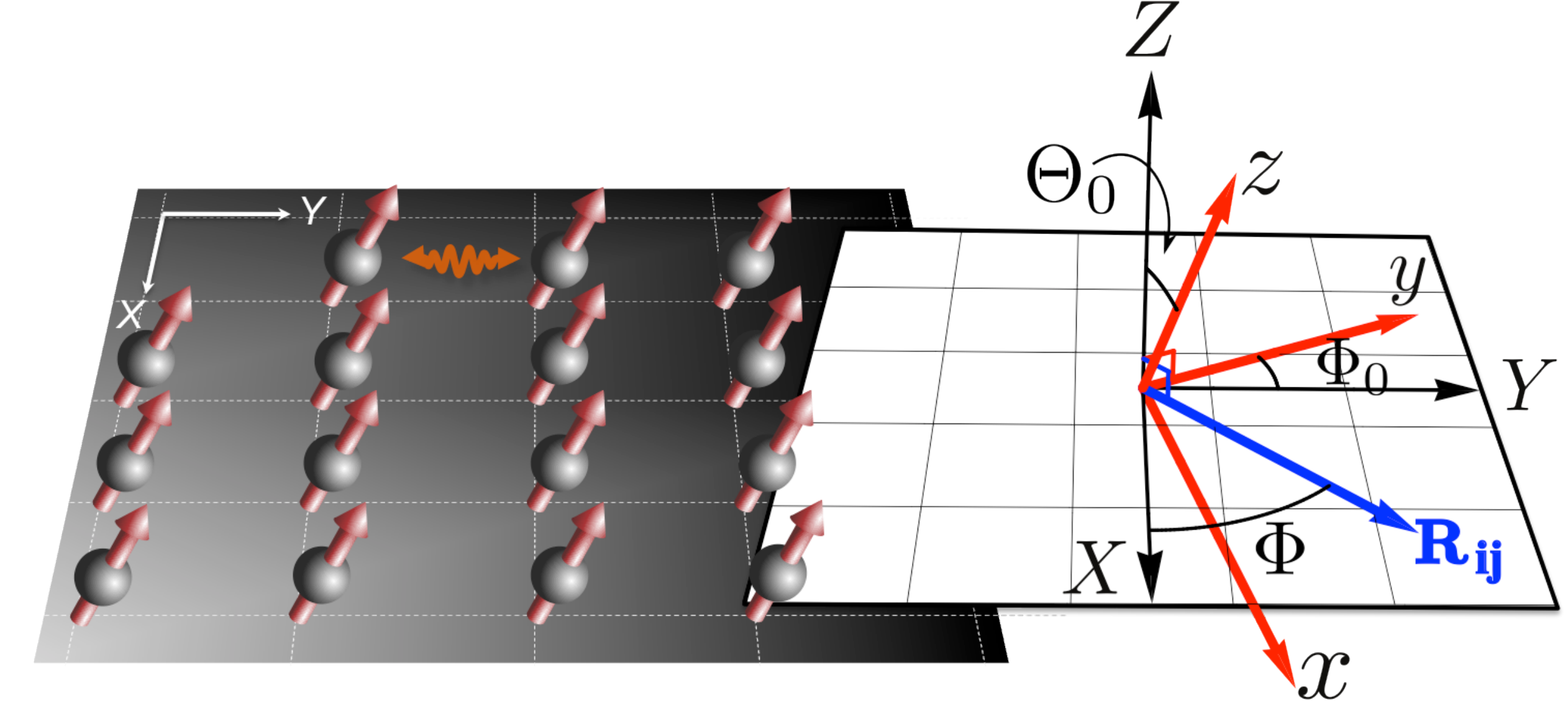}% Here is how to import EPS art
\caption{\label{schematic} 
Schematic representation of a 2D dipolar droplet.
The grey droplet represents a 2D array of interacting tilted dipoles. 
The dipoles are tilted by a static field in the $\hat{z}$ direction, oriented at $\Theta_0, \Phi_0$ relative to the lattice basis $\{X,Y,Z\}$. {\bf R}$_{ij}$ is a vector connecting dipoles in the XY plane.
 }
\end{figure}

%To be specific, we analyze our proposal in the context of ultra-cold polar molecules in an optical lattice \cite{Brown03, Gorshkov11,Micheli07,Aldegunde08}; the discussion naturally generalizes to any dipolar system which can exhibit a tripod-level structure (e.g. Rydberg atoms, NV defects, Dysprosium) \cite{Childress06,Pohl09,Tagliacozzo12,Lu10}. 

%TODO --- move exes above. 

Let us consider a square lattice composed  of fixed, three-state  magnetic or electric dipoles placed   in a static  external field. Such  an arrangement naturally arises in experimental systems ranging from ultra-cold polar molecules \cite{Micheli06,Micheli07,Aldegunde08,Gorshkov11,Chotia12,Aikawa10,Deiglmayr08} and Rydberg atoms \cite{Schempp10,Pritchard10,Tagliacozzo12} to solid-state spins \cite{Childress06,Balasubramanian09} and magnetic atoms \cite{Lu10}.  As shown in Fig.~\ref{schematic}, the dipoles occupy the $\{X,Y\}$ plane and couple via dipole-dipole interactions,
\begin{equation}
H_{dd} = \frac{1}{2} \sum_{i\neq j}  \frac{\kappa}{R_{ij}^3}   \left [  {\bf d}_i  \cdot {\bf d}_j - 3({\bf d}_i \cdot {\bf \hat{R}}_{ij})({\bf d}_j \cdot {\bf \hat{R}}_{ij}) \right ],
\end{equation} 
where $\kappa$ is $1/4\pi\epsilon_0$ for electric dipoles or $\mu_0 /4\pi$ for magnetic dipoles, and ${\bf R}_{ij}$ connects the dipoles ${\bf d}_i$ and ${\bf d}_j$. 
The three  states of each dipole, which we label as $|0\rangle$, $|\pm1\rangle$, are eigenvectors of the   $\hat{z}$-component of (rotational or spin) angular momentum. We assume that the $|\pm1\rangle$ states are degenerate while the $|0\rangle$ state is energetically separated from them (Fig.~2a). 

Each three-level dipole is driven by electromagnetic fields of Rabi frequency $\Omega_+$ (right-circularly polarized), $\Omega_-$ (left-circularly polarized) and detuning $\Delta$ as shown schematically in Fig.~2a. With $|\Omega_+|, |\Omega_-| \ll \Delta$, the approximate eigenstates (dressed states) are: 
$|0\rangle$, $|B\rangle = \alpha( |-1\rangle + \beta |1\rangle)$, and $|D\rangle =\alpha^{*}( -\beta^{*} |-1\rangle +  |1\rangle)$, where $\alpha = \Omega_+ / \tilde{\Omega}$, $\alpha\beta = \Omega_- / \tilde{\Omega}$, and  $\tilde{\Omega} = \sqrt{|\Omega_-|^2 + |\Omega_+|^2}$.  The energies of these dressed states are $E_0 =  -\tilde{\Omega}^2/\Delta$, $E_B = \Delta +\tilde{\Omega}^2/\Delta$, and $E_D = \Delta$ respectively.
We let $d$ represent the typical size of the dipole moment and $R_0$ be the nearest-neighbor spacing; by ensuring that $\kappa d^2/R_0^3 \ll \tilde{\Omega}^2/\Delta$ and so long as we initially avoid populating $|D\rangle$, the system remains within the subspace locally spanned by $|0\rangle$ and $|B\rangle$ (note that one could also choose to work in the subspace spanned by $|0\rangle$ and $|D\rangle$).

% TODO [discuss relationship to Dalibard]

%$C_q^k(\theta,\phi) = (\frac{4\pi}{2k+1})^{1/2} Y_{kq}(\theta,\phi)$ is proportional to the spherical harmonic and $T^2_q$ is the  second-rank tensor formed from the dipole operators on sites $i$ and $j$. 

% TODO Add discussion of energy conservation => B conservation

Thus, it is natural to view $|B\rangle$ as representing an effective hardcore bosonic excitation (spin-flip), while $|0\rangle$ represents the absence of such an excitation.  Recasting this system in terms of operators $a_i^{\dagger} = |B \rangle \langle 0 | _i$ ($n_i = a_i^{\dagger}a_i$) yields a 2D model of conserved hardcore lattice bosons, 
\begin{equation}
	\label{eq:ham_hcboson}
H_B =  -\sum_{ij} t_{ij} a_i^{\dagger} a_j + \frac{1}{2}\sum_{i \neq j} V_{ij} n_i n_j,
\end{equation}
where we define the hopping $t_{ij} = -\langle B_i 0_j | H_{dd} | 0_i B_j\rangle$, the on-site potential $t_{ii} = \sum_{j \neq i} ( \langle 0_i0_j | H_{dd}| 0_i0_j \rangle-\langle B_i0_j | H_{dd} | B_i0_j \rangle)$, and   the interaction $V_{ij} = \langle B_iB_j | H_{dd} | B_iB_j \rangle + \langle 0_i0_j | H_{dd}| 0_i0_j \rangle -  \langle B_i0_j | H_{dd} | B_i0_j \rangle - \langle 0_iB_j | H_{dd}| 0_iB_j \rangle$. 
The conservation of total boson number, $N_i = \sum_i a_i^{\dagger} a_i$, arises from the condition $\kappa d^2/R_0^3 \ll \Delta$, which ensures that particle-number non-conserving terms of $H_{dd}$ are energetically disallowed. 

\begin{figure}
\centering
\includegraphics[width=3.4in]{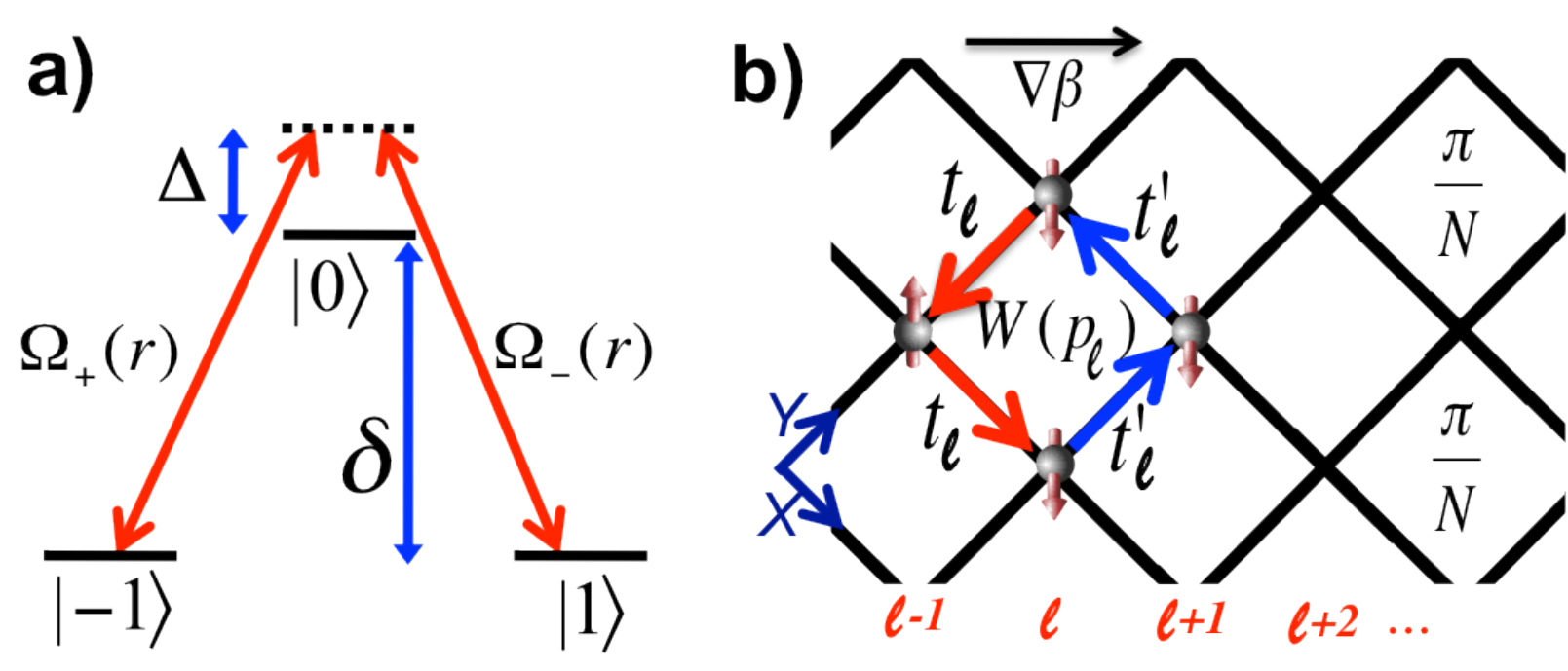}% Here is how to import EPS art
\caption{\label{fig:Fig2} 
a) Depicts the on-site level structure and the two-photon driving scheme. These levels could, for example, be adiabatically connected to the $J=1$ manifold of a rigid rotor as one  turns on a DC electric field (see Eq.~(6)).  The resonance frequency of the dressing lasers is detuned by $\Delta$, while their Rabi frequencies are $\Omega_{-}(r)$  and  $\Omega_{+}(r)$. We consider $|\Omega_{\pm}| \ll \Delta$ to operate in the far-detuned limit. In the case of polar molecules, $\delta$ is the electric-field induced splitting within the $J=1$ manifold, which we require to be larger than the typical dipolar interaction strength. 
b)  Square lattice with a single tilted dipole per vertex.   
We index columns of the lattice by $\ell$ and plaquettes by $p_\ell$. 
For a particle traversing the edge of a single plaquette, there are two contributions $t_\ell$ and $t_\ell'$ to $W(p_\ell)$; each contribution occurs twice as represented by the red and blue colored arrows. 
A simple periodic gradient of $\beta$ enables uniform $\pi/N$ flux per plaquette.   
}
\end{figure}

The functional form of the effective hardcore bosonic Hamiltonian  Eq.~(\ref{eq:ham_hcboson}) arises for any system of pinned, three-level dipoles. 
The parameters in $H_B$ are given by ($\kappa = 1, i \ne j$):
\begin{align}
	\label{eq:tij}
t_{ij} &=  \frac{d_{01}^2}{R^3} \left [ \chi_i^{\dagger} (q_0 + \text{Re}[q_2] \sigma^x + \text{Im}[q_2]\sigma^y) \chi_j \right ],  \\
t_{ii} &=  -\sum_{j \neq i} 2\frac{q_0}{R^3} (d^0 d^B_i - (d^{0})^2), \nonumber \\
V_{ij} &=  2\frac{q_0}{R^3} 
\left [ 
	d^B_i d^B_j 
	- d^0 d^B_i - d^0 d^B_j
	+ (d^0)^2
\right ], \nonumber
\end{align}
where
$d^{0}$ ($d^{B}$) is the permanent $\hat{z}$-dipole moment of the $|0\rangle$ ($|B\rangle$) state, 
$d_{01}$ is the transition dipole moment from $|1\rangle$ to $|0\rangle$ \cite{dplus}, 
$\chi_i = \alpha_i(1,\beta_i)^T$ is the normalized drive-spinor on site $i$, 
$q_0 = \frac{1}{2} (1-3\cos^2(\Phi-\Phi_0)\sin^2(\Theta_0))$,  
$q_2 = -\frac{3}{2} [ \cos (\Phi - \Phi_0) \cos \Theta_0 - i \sin (\Phi -\Phi_0) ]^2$, $\vec{\sigma}$ are the Pauli matrices, and   $(R,\Phi)$ is the separation ${\bf R}_{ij}$ in polar coordinates (Fig.~1).  We have suppressed the explicit $ij$ dependence of $R$, $\Phi$, $q_0$, and $q_2$. 
While the form of $d^B_i$, and hence of interactions, depends on the underlying implementation, 
the single-particle band structures that can be achieved via driving are independent of such details \cite{tii}.

Let us first explore these topological single-particle bands and  illustrate the interplay between the driven breaking of time-reversal and the anisotropic dipolar interaction. 
%
%by tuning the microwave driving schemes, on the other hand, are independent of such details and already reveal much interesting physics.
%
%
%The effective Hamiltonian, $H_B$, indeed provides a large parameter space for exploring the physics of bosons with controllable kinetic energies.
As a simple example, we demonstrate how to achieve a synthetic background gauge field with uniform flux $\pi/N$ per plaquette on a square lattice (assuming only nearest-neighbor hops). 
We choose the ``magic'' electric field tilt, 
$(\Theta_0, \Phi_0) = (\sin^{-1}(\sqrt{2/3}),\pi/4)$, 
where $q_0=0$ along $\hat{X}$ and $\hat{Y}$. This choice allows us to isolate the terms of $H_{dd}$ that harbor intrinsic phases, namely, those associated with $d_{i}^{+} d_{j}^{+}$ and $d_{i}^{-} d_{j}^{-}$, where $d_{\pm} = \mp (d_x \pm i d_y)/\sqrt{2}$ \cite{Gorshkov11,dplus}. Moreover, it simplifies the form of nearest-neighbor hopping to
% TODO check \omega\tilde
\begin{align}
t_{ij}^{\hat{X}} &=  \frac{d_{01}^2}{R_0^3}   \chi_i^{\dagger} \left [ \frac{1}{2} \sigma^x - \frac{\sqrt{3}}{2} \sigma^y \right ] \chi_j,  \nonumber \\ 
t_{ij}^{\hat{Y}} &=  \frac{d_{01}^2}{R_0^3}  \chi_i^{\dagger} \left [ \frac{1}{2} \sigma^x + \frac{\sqrt{3}}{2} \sigma^y \right ] \chi_j .
\end{align}
% TODO Use \vec{\Omega} notation?

% TODO Add diagonal hops (light dashed lines) to figure
\noindent The microscopic breaking of time-reversal arises from the asymmetry between left- and right- circularly polarized radiation and is captured by the ratio $\beta=\Omega_-/\Omega_+$. While each Rabi frequency is characterized by both an amplitude (intensity) and a phase, initially, we will consider only varying the amplitude of $\beta$; phase variations will be considered in more detail in the discussion of many-body states.  Physically, it is  $\beta$ which defines each hardcore boson $|B\rangle$, by setting the relative admixture between the $|1\rangle$  and $|-1\rangle$ states.  Keeping $\beta$ real, let us now consider varying the intensities of the drive fields along the $\Phi = \pi/4$ direction in a periodic fashion.

 %Let us consider a periodic drive pattern for $\beta \in \mathbb{R}$ (e.g. $\Omega_- , \Omega_+ \in \mathbb{R}$) along the $\Phi = \pi/4$ direction.  

 For each plaquette, we define the Wilson loop, $W(p) = \prod_{\partial p}t_{ij}$, which is identical along columns indexed by $\ell$ (Fig.~2b). The flux in a plaquette is then the phase of this Wilson loop, $\Psi_{\ell}=\arg[W(p_{\ell})]=\arg[t_\ell^2 t_\ell'^2]$, where $t_\ell$ are $t_\ell'$ are the hops depicted in Fig.~2b. Taking $\theta_\ell = \arg(t_\ell)$ and noting that $\theta_\ell' = \arg(t_\ell') = -\theta_{\ell+1}$ yields the phase of the Wilson loop as $\Psi_\ell = 2\theta_\ell-2\theta_{\ell+1}$. To achieve a uniform $\pi/N$  flux per plaquette, we can take $\theta_{\ell+1} = \eta - \ell \frac{\pi}{2N}$, where $\eta \in \mathbb{R}$ is a constant to be specified. 
% 
%We now prove for all $N \in \mathbb{Z}_{>0}$, that there exists a periodic choice of $\beta$ with a maximum periodicity of $4N$, which generates the desired uniform background flux. 
%
%
%Defining $r_\ell = \beta_{\ell+1}/\beta_{\ell}$ yields $\beta_{\ell+1}  = \prod_{k=1}^{\ell}r_k \beta_{1}$, implying that the periodicity of $\beta$ is the smallest $L$ such that $\prod_{k=1}^{L}r_k=1$. We now prove for all $N \in \mathbb{Z}_{>0}$, that there exists a periodic choice of $\beta$ with a maximum periodicity of $4N$, which generates the desired uniform background flux. First,  note that $t_\ell/\beta_{\ell-1} = r_\ell \bar{w} +w$ and hence, $\theta_\ell = \arg(r_\ell \bar{w} +w)$. Inspection reveals that one can then express $r_\ell$ as a function of $\theta_\ell$,
%
From the definition of $\theta_{\ell}$, one finds a simple recursion relation for $\beta$, 
\begin{equation}
\label{recursion}
\frac{\beta_{\ell+1}}{\beta_{\ell}} =\frac{\sin(\frac{\pi}{3}-\eta+\ell\frac{\pi}{2N})}{\sin(\frac{\pi}{3}+\eta-\ell\frac{\pi}{2N})},
\end{equation}
with maximum periodicity $4N$ \cite{supp}. Starting from any initial $\beta_1$, Eq.~(\ref{recursion}) yields a recursively generated drive pattern which achieves the desired uniform $\pi/N$ background gauge field. 

%For $N\in3\mathbb{Z}$, $\prod_{\ell=1}^{2N}\frac{\sin(\frac{\pi}{3}-\eta+\ell\frac{\pi}{2N})}{\sin(\frac{\pi}{3}+\eta-\ell\frac{\pi}{2N})}=1$ for all $\eta\in\mathbb{R}$, while for  $N\in \mathbb{Z} \backslash 3\mathbb{Z}$, $\prod_{\ell=1}^{4N}\frac{\sin(\frac{\pi}{3}-\eta+\ell\frac{\pi}{2N})}{\sin(\frac{\pi}{3}+\eta-\ell\frac{\pi}{2N})}=1$ for $\eta=\lambda\pi/N$ with integer $\lambda$. Starting from any initial $\beta_1$, Eq.~(\ref{recursion}) yields a recursively generated drive pattern which achieves the desired uniform $\pi/N$ background gauge field. 

%Note that, in both cases, $\eta$ must be chosen so that the denominator of the product is never zero. For each choice of $\pi/N$, the above allows one to recursively generate an infinite number of periodic $\beta$-microwave drive-patterns, which induce the desired uniform background gauge field. 

While the uniform flux per plaquette is reminiscent of the square lattice Hofstadter problem   \cite{Hofstadter76}, we emphasize that the physics of these driven dipoles is significantly richer, owing to the additional modulation of $t_{ij}$.   The background flux field arises, in part, from the natural phases associated with the dipolar interaction. This ensures that  (as in \cite{Cooper11}) the number of flux quanta per plaquette is not limited by the magnitude of laser intensities,  contrasting with the majority of previous synthetic gauge field proposals, where the scaling to high artificial fluxes is extremely difficult \cite{Lin09, Dalibard11,Juzeliunas06,Abo-Shaeer01, Spielman09}. 

% TODO Perhaps rather comment that while this is 'uniform flux' it is nonuniform hopping and it is not simply the square lattice Hofstadter problem because of diagonal and long-range hops.
%\paragraph{Two-site unit cell.} 

%Remaining at the ``magic'' tilt, we now turn to a detailed study of $H_B$ restricted to a two-site unit cell, as depicted in Fig.~3a.

To illustrate the symmetry breaking required for the generation of gapped Chern bands, we now  turn to a detailed study of $H_B$ restricted to a two-site unit cell (remaining at the ``magic'' tilt), as depicted in Fig.~3a.  This restriction has the virtue of being analytically tractable and allows us to identify the anti-unitary symmetries associated with the Dirac points \cite{Thouless82, Hatsugai93}. Let us consider  $\beta = \beta_1,\beta_2$ on the two sites of the unit cell and include all terms up to next-next-nearest neighbor. The topology of the bands depends on the relative ratio of $\beta_1$ and $\beta_2$. For $\beta_1 \in \mathbb{R}$, the phase diagram in Fig.~3b illustrates the Chern invariant of the bottom band as a function of the complex $\beta_2$-plane. There exist two circles of gapless (Dirac) points protected by distinct anti-unitary symmetries. 

%On the outer circle, the Hamiltonian harbors an effective time-reversal symmetry evidenced by the fact that all Wilson loops are real. 
%On the inner circle, the Hamiltonian is invariant under conjugation combined with inversion through the mid-point of an edge. 
%Away from the magic tilt and including longer range terms, both circles  deform smoothly. For generic $\beta_1$ and $\beta_2$, these Dirac points gap in a manner that causes the bands to become topologically non-trivial (Fig.~3b). Underlying these non-trivial Chern bands is the aforementioned synthetic gauge field associated with the effective dressed spin-flips. 

\begin{figure}
\centering
\includegraphics[width=3.4in]{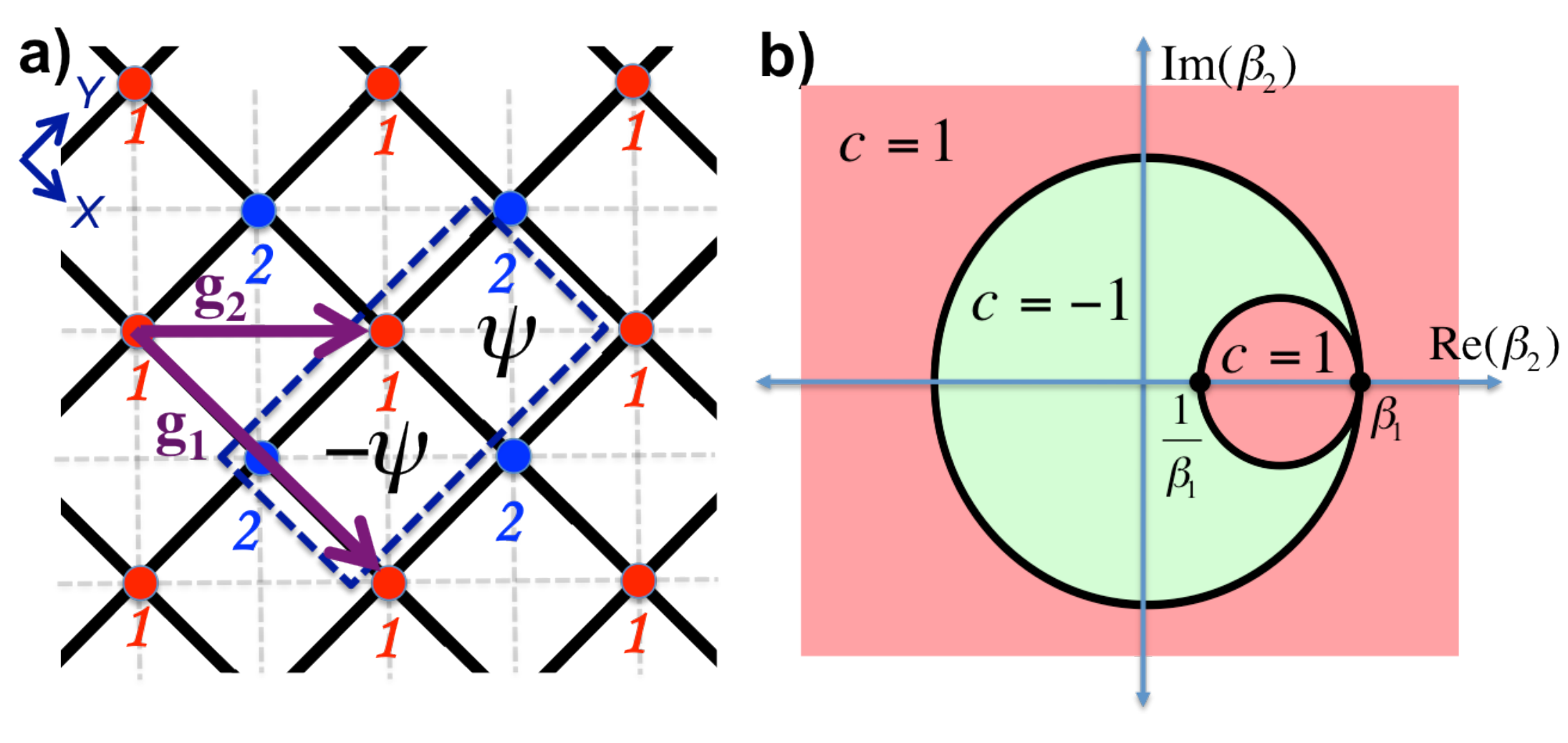}
\caption{\label{fig:twosite} 
a) Schematic representation of the two-site unit cell lattice with $\beta = \beta_1,\beta_2$. 
The dotted box outlines a single unit cell. 
There is a flux $\Psi, -\Psi$ which alternates in neighboring square plaquettes. The direct lattice vectors $g_1$ and $g_2$ are depicted as purple arrows. While all hops are present with amplitude decaying as 1$/R^3$, only nearest-neighbor (solid) and next-nearest-neighbor (dashed) hops are shown.
b) The topology of bulk bands as a function of complex $\beta_2$ for $\beta_1 \in \mathbb{R}$. The Chern number is $c= \frac{1}{4\pi} \int dk_x dk_y (\partial_{k_x} \hat{d} \times \partial_{k_y} \hat{d})\cdot \hat{d}$, where $H(k) = \vec{d}(k)\cdot\vec{\sigma} + f(k) $.  }
\end{figure}

\emph{Implementation}---An experimental realization of our proposal can be envisioned with either electric (e.g.~polar molecules) or magnetic (e.g.~solid-state spins) dipoles.   As previously mentioned, the form of $d_i^B$ depends on this choice, since the permanent dipole moment of the $|\pm 1\rangle$ states have either the same or opposite signs.  We emphasize that the long intrinsic lifetimes of such systems make them ideal for the consideration of driven, non-equilibrium phenomena \cite{Brown03,Maurer12}.

%mws polarized in little x y. 

To be specific, we now focus on diatomic polar molecules (trapped in a deep optical lattice) in their electronic and vibrational ground state. We utilize microwave fields to dress the molecules and partially polarize them with an applied DC electric field along $\hat{z}$ (Fig.~1); ignoring electronic and nuclear spins, this yields a single-molecule Hamiltonian, 
\begin{equation}
H_m= B J^2  - d_z E   + H_{D}, 
\end{equation} 
where $B$ is the rotational constant, $J$ is the rotational angular momentum operator, $d_z$ is the $\hat{z}$ component of the dipole operator, $E$ is the magnitude of the applied DC field, and $H_{D}$ characterizes the  dressing of the $J=1$ rotational states depicted in Fig.~2a \cite{Brown03, Gorshkov11}.

In the absence of applied fields, each molecule possesses rigid rotor eigenstates $|J,M\rangle$.  The applied electric field $\bf{E}$ mixes eigenstates with the same $M$,  splitting the degeneracy within each $J$ manifold and inducing a finite permanent dipole moment for each perturbed rotational state.  
We choose from among these states to form the effective three-level dipole; an example of one possibility for $|0\rangle$, $|\pm 1\rangle$ is shown in  Fig.~2a. Since these $|\pm 1\rangle$ states have an identical induced dipole moment $d^1$, one finds that $d^{B}_i = d^{1}$, and hence,
\begin{eqnarray}
	\label{eq:tijPM}
V_{ij} &=&  2\frac{q_0}{R^3} (d^{0} - d^{1})^2.
\end{eqnarray}
%Owing to the lattice geometry, the potential $t_{ii}$ for each site is identical and hereon, we will drop this overall shift.  
The relative strength of the interaction $V_{ij} / t_{ij}$ is thus set by $(d^{0} - d^{1})^2/d_{01}^2$;  this is a highly tunable parameter and can easily reach $\sim 100$ for certain choices of rotational states and DC electric field strengths \cite{Gorshkov11}. 

%This tunability suggests the possibility of observing interaction-dominated many-body phases in such a polar-molecule implementation. 

The main challenge in an experimental realization of our proposal lies in the spatial modulation of the drive fields at lattice scale. For spins in the solid-state and on-chip polar molecule experiments, one might envision using near-field  techniques.  A more straightforward approach, suitable for molecules,  is to utilize pairs of optical Raman beams (see supplementary information for details) \cite{Yao12b}.  For example, the so-called lin$\perp$lin configuration  \cite{Dalibard89} automatically ensures that $\tilde{\Omega}$ and $\Delta$ are identical on all sites and moreover, generically produces gapped topological band-structures.  

\begin{figure}
\centering
\includegraphics[width=3.4in]{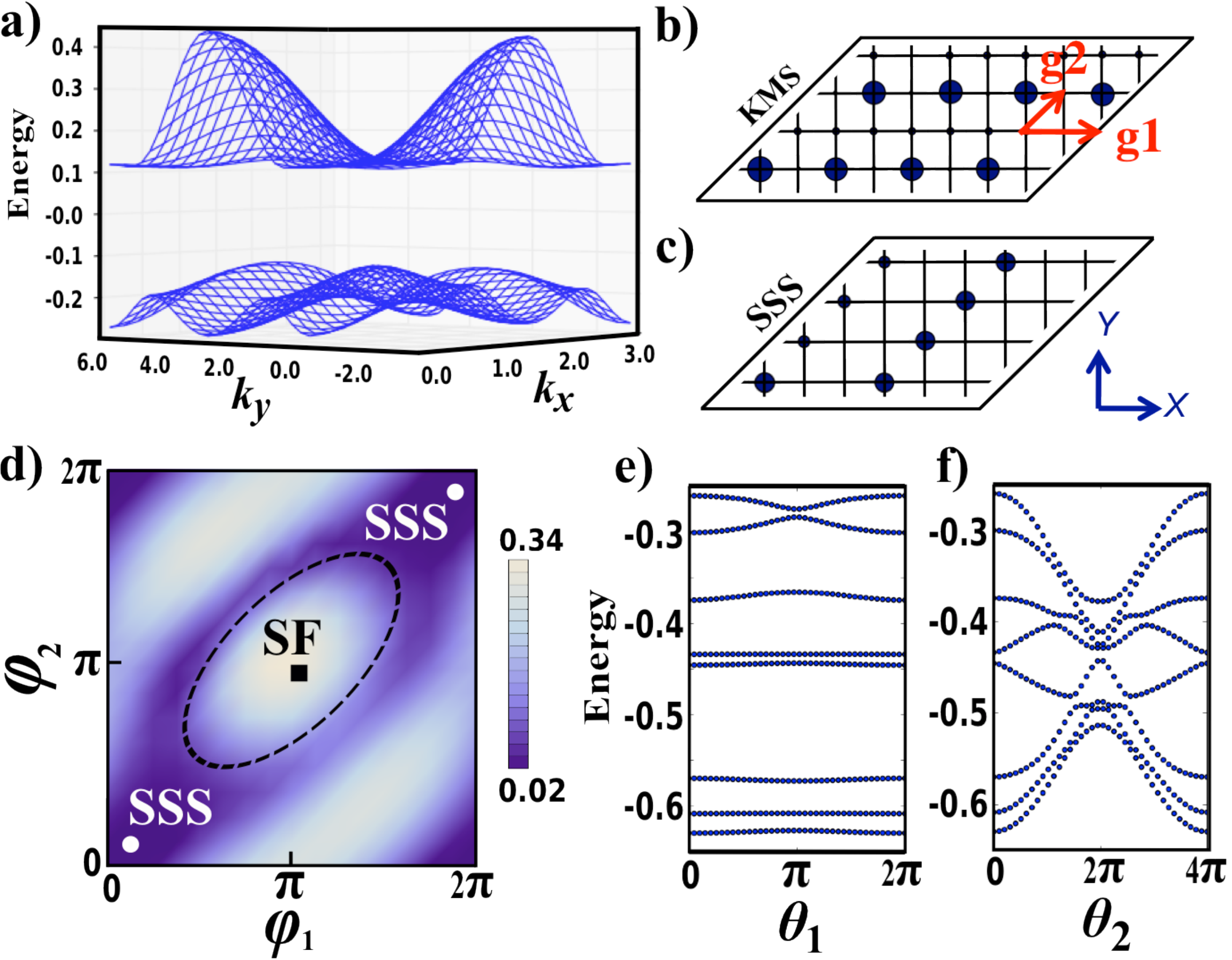}
\caption{\label{fig:manybody} 
Phase Transitions in topological flat bands of 2D driven dipoles.
a) Band structure for $(\Theta_0,\Phi_0) = (0.46,0.42)$, $\beta_1 = 3.6 e^{2.69i}$, and $\beta_2 = 5.8 e^{5.63i}$. We have verified that the Chern number does not change upon adding in dipolar interactions up to order $1/27R_0$. Significantly flatter band structures with flatness ratio $>10$ can be obtained for slightly generalized  configurations involving a tripod level-structure and optical super-lattice \cite{Yao12b}.
b) Structure factor $S(R,0) = \langle n(R)n(0)\rangle$ for filling $\nu=1/2$ in KMS \cite{supp} and c) SSS regime; size of circles indicates weight. 
d) Spectral gap density plot as a function of varying MW drive for parameters: $(\Theta_0,\Phi_0) = (0.66,\pi/4)$, $\beta_1 = -2.82 e^{i \phi_{1}}$,  $\beta_2 = -4.84 e^{-i\phi_{2}}$ and $(d^{0}-d^{1})^2/d_{01}^2 \approx 2.8$. 
The transition from the SF, which has a unique finite-size ground state, to the degenerate SSS shows as a collapse of this gap.
e) Spectral flow in the ground state momentum sector of the SSS under twisting of the boson boundary condition in the $\hat{g}_1$ and f) $\hat{g}_2$ directions.
For the $N_s=24$ lattice with 6 bosons, momentum sectors return to themselves after $2\pi$ in $\theta_1$ and after $4\pi$ in $\theta_2$.
} 
\end{figure}

%Here, we explore the ground state phases of the effective Hamiltonian \eqref{eq:ham_hcboson} in the two-site driving scheme and leave the many interesting dynamical questions of this driven system to future work. 

%We  leave aside thermodynamic questions regarding the long dipolar tail and simply truncate all terms at the next-next-nearest neighbor (NNNN) level.

%, which for hardcore bosons is approximately given by the band gap.

\emph{Many-body phases}---To illustrate the power of the present approach, we briefly explore two examples of correlated ground state phases which arise in the Hamiltonian Eq.~(2).
As $H_B$ conserves boson number $N$, we may  consider its many-body physics at finite filling fractions $\nu$ (particle number per unit cell). 
Let us work with a two-site unit cell and truncate the dipolar interactions  at next-next-nearest-neighbor  order. 
Bosons residing in a strongly dispersing band structure generically form superfluids in order to minimize their kinetic energy. 
Interaction dominated phases arise when the single-particle bands disperse less than the scale of interactions. 
Numerical optimization of the flatness ratio (bandgap/lowest bandwidth) over the six-dimensional parameter space of microwave driving and tilt angle reveals approximately flat Chern bands in several regions of phase space. The flatness of these bands (Fig.~4a) derives from interference between the  hopping in different directions and, microscopically, owes to an interplay between the natural anisotropy associated with dipolar interactions and the spatial variation of the drive fields. 
%These quenched band structures provide a fertile starting point for discovering strongly correlated phases \cite{Sun11, Tang11,Neupert11, Trescher12,Yang12, Wang12}. 

%We study lattices of size: $N_s=24$ ($3\times4\times2$), $N_s=32$ ($4\times4\times2$), and $N_s=40$ ($5\times4\times2$).

% Example 1

As a first example, we consider the band structure depicted in Fig.~4a, where the lower/upper band carry Chern index, $c = \mp1$ (parameters in caption). 
Exact diagonalization at filling fraction $\nu=1/2$ and relative interaction strength $(d^0 - d^1)^2/d_{01}^2 \approx 6$ reveals a knight's move solid (KMS) phase with a 4-fold degenerate, gapped, ground state. 
The real-space structure factor $S(R,0) = \langle n(R) n(0) \rangle$ (at total number of sites, $N_s=32$) in Fig.~4b illustrates the knight's move relationship of the bosons in the ground state.
Twisting the boundary condition of the KMS in the $\hat{g}_1$, $\hat{g}_2$ directions (Fig.~4b) does not significantly affect the ground state energy, as expected of an insulator \cite{supp}.

Many other commensurate phases arise as as we tune the driving fields to other regions of phase space. Figure 4d shows a phase diagram containing both superfluid (SF) and striped supersolid (SSS) phases.
We can characterize the SSS arising at $\phi_{1} = \phi_{2} = 0.1$ as follows:
First, diagonalization reveals the existence of three degenerate ground states in the sectors: $k_2=0, k_1 = 0, 2\pi/3, 4\pi/3$.
Consistent with striped ordering, the structure factor shows density stripes in the $\hat{g}_2$ direction (Fig.~4c). 
However, each of these stripes has incommensurate boson number, suggesting  delocalization along the stripes.
To wit, for $N_s=24$, the 6 hardcore bosons are distributed evenly along two stripes, each containing 4 sites. 
Strong phase coherence along the stripes shows up in the sensitivity to twists in the $\hat{g}_2$ direction, while transverse twists produce essentially no dispersion, as shown in Fig.~\ref{fig:manybody}e,f.

%Enlarging the system to $N_s=36$ does not alter this qualitative picture.  

%We note that the stripe ordering breaks the underlying translational symmetry of the lattice but not any existing rotational symmetry, as these are explicitly broken by the tilt of the dipoles.

%The above considerations indicate  that driven  dipolar spin arrays can effectively generate topological nearly flat bands.  Implementations using  either diatomic polar molecules in optical lattices \cite{Micheli06,Micheli07,Aldegunde08,Gorshkov11,Chotia12,Aikawa10,Deiglmayr08} or localized spins in the solid-state (such as those associated with NV defect centers \cite{Childress06}) 
%naturally yield examples of interaction-dominated phases. 

\emph{Conclusion}---Our proposal  opens the door to a number of intriguing directions. In particular, the adiabatic preparation and detection of single-excitation states may provide an elegant approach to probing chiral dynamics, edge modes, and the Chern index \cite{Yao12,Yao12b, Abanin12}. More generally, dynamical preparation, manipulation and detection of many-body states in such driven topological systems remains an exciting open question \cite{Sorensen10}. Finally, the large available parameter space holds the promise of more exotic phases, such as fractional Chern insulators  \cite{Yao12b}. Realizing such phases in an effective  spin system may provide a deeper understanding of the stability of such states in the context of generalized long-range dipolar interactions.

\emph{Acknowledgements}---We gratefully acknowledge conversations with A. Zhai, B. Lev, J. Preskill, J. Alicea, and N. Lindner. This work was supported, in part, by the NSF, DOE (FG02-97ER25308), CUA, DARPA, AFOSR MURI, NIST, Lawrence Golub Fellowship, Lee A. DuBridge Foundation, IQIM and the Gordon and Betty Moore Foundation.


\begin{thebibliography}{99}
\bibitem{tii} The potential $t_{ii}$ for each site is independent of $i$ for polar molecules on a square lattice (since $d^{B}_i = d^1$) and for NV centers generally (since $d^0=0$).  Alternatively, such an on-site potential can  be corrected using  a spatially dependent Stark shift.
\end{thebibliography}
\end{document}

%\preprint{APS/123-QED}

\begin{center}
\begin{large}

\bf{Supplemental Material: Topological flat bands from two-dimensional dipoles}
\end{large}

\end{center}

\vspace{5mm}


\noindent {\bf Proof of uniform $\pi/N$:} Here, we prove that a periodic drive pattern enables arbitrary uniform $\pi/N$ flux per square plaquette, as shown in Fig.~2b of the main text.  The flux in each plaquette $p$ is the phase of the Wilson loop, $W(p) = \prod_{\partial p}t_{ij}$.    For a given plaquette, $p_\ell$, $\Phi_{\ell} = \arg[W(p_\ell)] = \arg[t_\ell^2 t_\ell'^2]$, where $t_\ell = \beta_{\ell -1} w + \beta_{\ell} \bar{w}$, $t_\ell' = \beta_{\ell} \bar{w} + \beta_{\ell+1} w$ (dropping the normalization factor which will not contribute to the argument) and $w=e^{i\pi/3}$. Taking $\theta_\ell = \arg(t_\ell)$ and $\theta_\ell' = \arg(t_\ell')$ yields the phase of the Wilson loop as $\Phi_\ell = 2\theta_\ell+2\theta_\ell' = 2\theta_\ell-2\theta_{\ell+1} $.  To achieve a uniform $\pi/N$  flux per plaquette, we can take $\theta_{\ell+1} = \eta - \ell \frac{\pi}{2N} + a_\ell \pi$ (note that in the main text the $a_\ell$ term was dropped for notational simplicity), where $a_{\ell} \in \{0,1\}$ and $\eta \in \mathbb{R}$ is a constant to be specified.  Defining $r_\ell = \beta_{\ell+1}/\beta_{\ell}$ yields $\beta_{\ell}  = \prod_{k=1}^{\ell-1}r_k \beta_{1}$, implying that the periodicity of $\beta$ is $L$ such that $\prod_{k=1}^{L}r_k=1$. We now prove, for all $N \in \mathbb{Z}_{>0}$, that there exists a periodic choice of $\beta$ with a maximum periodicity of $4N$ which generates the desired uniform background flux. First,  note that $t_{\ell+1}/\beta_{\ell} = r_{\ell} \bar{w} +w$ and hence, $\theta_{\ell+1} = \arg(r_{\ell} \bar{w} +w) + b_{\ell} \pi$, where $b_{\ell} \in \{0,1\}$. Simple geometry reveals that taking
\begin{equation}
\label{recursion}
r_{\ell}=\frac{\beta_{\ell+1}}{\beta_{\ell}} =\frac{\sin(\frac{\pi}{3}-\eta+\ell\frac{\pi}{2N})}{\sin(\frac{\pi}{3}+\eta-\ell\frac{\pi}{2N})}
\end{equation}
ensures that $\arg(r_{\ell} \bar{w} +w) = \eta-\ell\frac{\pi}{2N} + c_{\ell} \pi$, where $c_{\ell} \in \{0,1\}$. Thus, we have succeeded in achieving $\theta_\ell$ of the desired form with $a_{\ell} = b_{\ell} + c_{\ell} \text{(mod 2)}$. Inspection reveals that for $N\in3\mathbb{Z}$, $\prod_{\ell=1}^{2N}\frac{\sin(\frac{\pi}{3}-\eta+\ell\frac{\pi}{2N})}{\sin(\frac{\pi}{3}+\eta-\ell\frac{\pi}{2N})}=1$ for all $\eta\in\mathbb{R}$, while for  $N\in \mathbb{Z} \backslash 3\mathbb{Z}$, $\prod_{\ell=1}^{4N}\frac{\sin(\frac{\pi}{3}-\eta+\ell\frac{\pi}{2N})}{\sin(\frac{\pi}{3}+\eta-\ell\frac{\pi}{2N})}=1$ for $\eta=\lambda\pi/N$ with integer $\lambda$.~\noindent~\QEDB

\begin{figure}[t]
  \begin{center}
   \includegraphics[width=5.4in]{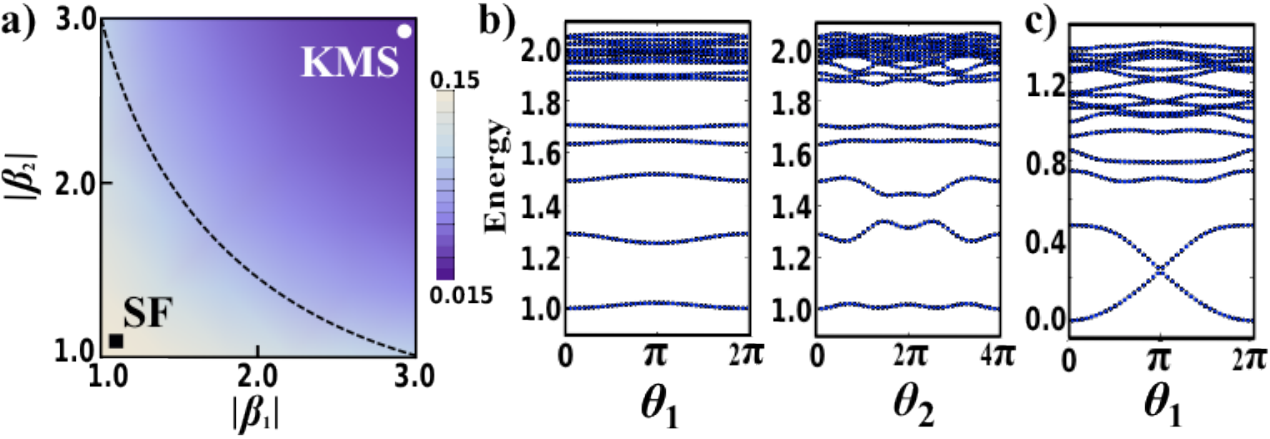}
  \end{center}
    \begin{flushleft}
\noindent \singlespacing FIG.~S1: \textbf{Knight's move solid.} a) Exact diagonalization at $N_s = 24$ reveals a spectral gap density plot, which varies as a function of MW drive amplitude for $(\Theta_0,\Phi_0) = (0.21,0.42)$, $\beta_1 = 2.8 e^{1.69i}$, and $\beta_2 = 2.8 e^{5.63i}$. Although the MW parameters are slightly different than in the main text, the qualitative KMS phase is unchanged; this choice of parameters better illustrates the transition from a SF, which has a unique finite-size ground state, to the degenerate KMS. b) Spectral flow of the KMS ground state in the $k_1=k_2=0$ momentum sector under twisting the boson boundary condition in the $\hat{g}_1$ and $\hat{g}_2$ directions for parameters indicated by the white circle in a). For the $N_s=24$ ($3\times4\times2$) lattice with 6 bosons, momentum sectors return to themselves after $2\pi$ in $\theta_1$ and after $4\pi$ in $\theta_2$. In the KMS phase, the ground state energy is insensitive to twisting. (c) This contrasts with the quadratic response of the energy in the phase-coherent SF. Depicted is the spectral flow of the $k_1=k_2=0$ momentum sector under twisting in the $\hat{g}_1$ direction for parameters indicated by the black square in a).  
\end{flushleft}
\end{figure}

As a simple example, to generate a uniform $\pi/3$ background gauge field, one can utilize a drive pattern with period $2N=6$; with $\beta = \{1,5,-10,\frac{5}{2},\frac{5}{6}, \frac{2}{3} \}$ (e.g. $\eta = \tan^{-1}(\frac{1}{3\sqrt{3}})-\pi/6$), one finds that the topological bandstructure exhibits six bands, where the lowest carries Chern number $c = -1$. We have verified that adding in further neighbor interactions (up to $1/27R_0$) does not alter the Chern number of the lowest band. While such additional interactions do indeed lead to a change in the dispersion of the band structure and hence the flatness, we find that (for the two-band models considered in the maintext) re-optimizing at each order always enables us to recover the previous flatness.

%Certain configurations yield bands with $c>1$ and and may provide a promising framework to search for topological superfluids. 

\vspace{10mm}
\noindent {\bf Knight's Move Solid:} The knight's move solid discussed in the maintext exhibits 4 degenerate ground states in  momentum sectors $k_1=0, k_2=0, \pi/2, \pi, 3\pi/2$. As shown in Fig.~S1(b), twisting the boundary condition in the $\hat{g}_1$, $\hat{g}_2$ directions (Fig.~4b) does not affect the ground state energy consistent with an insulator.  By enlarging the system size, we confirm that the finite-size excitation gap of the KMS appears to remain finite in the thermodynamic limit.

% TODO Translate this 'amplitude' into the spinors
As we vary the magnitude of $\beta_1$ and $\beta_2$, keeping all other parameters fixed, the spectral gap $E_1 - E_0$ changes dramatically (Fig.~S1(a)), opening up as either amplitude approaches 1 (corresponding to linearly polarized MW driving). 
At $|\beta_1| = |\beta_2| = 1$, diagonalization reveals a unique ground state, residing in the $k_1 = k_2 = 0$ sector. 
While our system sizes are too small to clearly observe the Goldstone mode of a superfluid (SF), in contrast to the KMS, twisting the boundary conditions dramatically alters the ground state energy, as depicted in Fig.~S1(c); this is consistent with a SF which harbors long-range phase coherence.

\vspace{10mm}
\noindent {\bf Optical Raman Dressing:} Our proposal can also be carried out in currently available ultracold polar molecules, such as  ${}^{40}$K${}^{87}$Rb, ${}^7$Li${}^{133}$Cs, and ${}^{41}$K${}^{87}$Rb. Spatial modulation of the bright state can be achieved by utilizing optical Raman beams to dress the rotational states of each molecule. For the upper leg of the Raman transition, we propose to utilize the $|J',m'\rangle = |2, 2\rangle$ rotational states of the $v' = 41$ vibrational level of the ${(3)}^1\Sigma^+$ electronic state. These states harbor  a strong 640 nm transition to the ground state and enable modulation of the bright state at lattice length scales.

%, opening up as either amplitude approaches 1 (corresponding to linearly polarized MW driving). 
%While our system sizes are too small to clearly observe the Goldstone mode of a superfluid (SF), 

% TODO Translate this 'amplitude' into the spinors